\documentstyle[prl,aps,preprint,epsf]{revtex}
\begin{document}
\newcommand{\be}{\begin{equation}}
\newcommand{\ee}{\end{equation}}
\newcommand{\bq}{\begin{eqnarray}}
\newcommand{\eq}{\end{eqnarray}}
\newcommand{\Sc}{Schr\"odinger\,\,}
\newcommand{\Sp}{\,\,\,\,\,\,\,\,\,\,\,\,\,}
\newcommand{\Ssp}{\,\,}
\newcommand{\no}{\nonumber\\}
\newcommand{\tr}{\text{tr}}
\newcommand{\Tr}{\text{Tr}}
\newcommand{\p}{\partial}
\newcommand{\la}{\lambda}
\newcommand{\La}{\Lambda}
\newcommand{\G}{{\cal G}}
\newcommand{\D}{{\cal D}}
\newcommand{\LL}{{\cal L}}
\newcommand{\OO}{{\cal O}}
\newcommand{\E}{{\cal E}}
\newcommand{\W}{{\bf W}}
\newcommand{\HH}{{\cal H}}
\newcommand{\de}{\delta}
\newcommand{\al}{\alpha}
\newcommand{\bi}{\beta}
\newcommand{\ga}{\gamma}
\newcommand{\Ga}{\Gamma}
\newcommand{\ep}{\epsilon}
\newcommand{\vep}{\varepsilon}
\newcommand{\th}{\theta}
\newcommand{\om}{\omega}
\newcommand{\J}{{\cal J}}
\newcommand{\pr}{\prime}
\newcommand{\ka}{\kappa}
\newcommand{\si}{\sigma}

\setcounter{page}{0}
\def\footnoterule{\kern-3pt \hrule width\hsize \kern3pt}
\tighten
\title{
Schr\"odinger Representation of $CP(N)$ Model for Large $N$ }
\author{Phillial Oh
\footnote{Permanent Address:
Department of Physics, Sung Kyun Kwan University,
Suwon 440-746, Korea;
Email address: {\tt ploh@ctpa03.mit.edu}}
and Jiannis Pachos
\footnote{Email address: {\tt pachos@ctp.mit.edu}}
}
\address{Center for Theoretical Physics \\
Massachusetts Institute of Technology \\
Cambridge, Massachusetts 02139 \\
{~}}

\date{MIT-CTP-2759,~~ July 1998}
\maketitle

\thispagestyle{empty}

\begin{abstract}
We examine the 1+1 dimensional $CP(N)$ model in the large $N$ limit
by using the \Sc representation. Starting from  the Hamiltonian analysis
of the model, we present the variational gap equation resulting from
the Gaussian trial wave functional.  The renormalization of the theory is 
performed with  insertion of mass and energy  counter-terms, and the 
dynamical generation of  mass and the energy eigenvalue are derived.
\end{abstract}

\vspace*{\fill}

\newpage

\section{Introduction}

The \Sc representation \cite{Symanzik,Jackiw}
 has been proven to be a powerful method for probing the 
non-perturbative aspects of quantum field theory 
\cite{Mansfield,Nair,Kim,MSP}. 
It provides a natural way of extracting physical quantities 
as the eigenvalues of the equivalent operators acting on wave functionals, 
which describe the theory. An effective way of computing the Hamiltonian 
eigenvalue is to use the variational method \cite{jack1,Pi,Yee}. 
Especially, the Gaussian trial wave functional has been extensively
used in a variety of models, and it reproduces the one-loop results 
 \cite{stev}.

In this paper, we reexamine the 1+1 dimensional $CP(N)$ model \cite{Eich}
for large $N$ \cite{Dada,coleman}. The purpose is to test 
the Gaussian approximation in a geometric theory with non-polynomial 
interactions, which is known to be exactly soluble in the large $N$ limit. 
The study also gives some insight into gauge theories
where similar features such as dynamical mass generation
and asymptotic freedom  arise. 
We approximate the vacuum wave functional with a Gaussian and  calculate 
the energy eigenvalue by minimizing the expectation value of the Hamiltonian.
The ensuing gap equation can be solved with a mass counter-term, and 
the solution exhibits the asymptotic freedom as expected. 
The renormalization  of the Hamiltonian is performed with the insertion of 
an extra energy  counter-term and the result is in agreement with the path 
integral treatment \cite{coleman}.

\section{Large $N$ $CP(N)$ model} 

We first use the coadjoint orbit method to formulate the $CP(N)$
model  \cite{oh1} in the large $N$ limit. Let us introduce
\be
Q=gKg^{-1},
\ee
where $g \in G=SU(N+1)$ and $K= {i \over N} \text{diag}(N,-1,-1,..,-1)$. 
The action of the $CP(N)$ model is given by
\be
S=-{1 \over 2\la} \int d^2x \Tr (\nabla _{\mu} Q \nabla^{\mu} Q),
\label{action1}
\ee
while $\la$ is  coupling constant, and 
$\nabla_\mu=\partial/\partial x^\mu$.
Parameterizing $g$ by $(N+1)$ vectors as $g=(\vec{Z}_1, \vec{Z}_2,..,
\vec{Z}_{N+1})$ 
with $\vec{Z}_p \in {\bf C}^{N+1}$ ($p=1,..,N+1$), such that 
$\vec{Z}_p^* \cdot \vec{Z}_q = \de _{pq}$, 
and $\det(\vec{Z}_1,\vec{Z}_2,..,\vec{Z}_{N+1})=1$, we obtain
\be
S={1 \over \la} \int d^2x (\nabla _{\mu} 
\vec{Z}_1^* \cdot\nabla ^{\mu} \vec{Z}_1 +
(\vec{Z}_1^* \cdot \nabla_\mu \vec{Z}_1)^2),
\label{act}
\ee
with the constraint $ {\vec{Z}_1}^* \cdot \vec{Z}_1=1$, 
for $\vec{Z}_1=(z_1, \cdots, z_{N+1})$. 
This constraint can be solved by assuming $z_{N+1}=z^*_{N+1}$. 
Introducing the Fubini-Study coordinate:
\be
\psi^a =\sqrt{N} {z_a \over z_{N+1}} \,\, ,
\ee
we obtain
\be
z_a={ \psi^a \over \sqrt{N+|\psi|^2}}\,\, , 
\Sp z_{N+1}={\sqrt{N} \over \sqrt{N+|\psi|^2}} \,\, .
\ee
Substitution  into (\ref{act}) gives 
\be
S= {1 \over \la} \int d^2 x g_{ab}\p_\mu \psi^a 
\p ^\mu \bar \psi ^b \,\, ,
\label{actnow}
\ee
where $g_{ab}$ is the standard Fubini-Study metric on $CP(N)$:
\be
g_{ab}={ (N+|\psi|^2 ) \de_{ab} - \bar \psi^a \psi^b 
\over (N+|\psi|^2)^2} \,\, , \Sp|\psi|^2=|\psi_1|^2+\cdots
+|\psi_N|^2 \,\, ,
\label{fubini}
\ee
where $\psi^a$ is an unconstrained $N$-component field. 
Note that $g_{ab}$ can be obtained from the K\"ahler potential
$K(\psi, \bar \psi)=\ln (N+|\psi|^2)$ by
\begin{equation}
g_{ab}={ \partial^2 K(\psi, \bar\psi)\over 
\partial\psi^a\partial\bar\psi^b}.
\end{equation}

To perform Hamiltonian analysis, we introduce the conjugate momenta, 
$\pi_a$ and $\bar \pi_a$ of $\psi^a$ and $\bar \psi^a$ respectively by 
\be
\pi_a = {1 \over \la}g_{ab}\dot{\bar \psi^b} \,\, , 
\Sp\bar \pi_a= {1 \over \la} g_{ba} \dot{\psi}^b \,\, .
\ee
Then the Hamiltonian density $\HH$ is given by 
\be
\HH= \pi_a \dot{ \psi}^a + \bar \pi _a \dot{\bar \psi^a} -\LL = 
\la g^{ab} \bar \pi_a \pi _b + {1 \over \la} g_{ab} \nabla \psi^a 
\nabla \bar \psi^b \,\, ,
\ee
where $g^{ab}$ is the inverse of $g_{ab}$:
\be
g^{ab}={1 \over N} (N+|\psi|^2)(N\de_{ab}+\bar \psi^a \psi^b) \,\, .
\ee
The canonical quantization
gives  the following equal time commutators:
\bq
&&
[\psi^a (x),\pi_b(y)]=i \hbar \de^a_{~b}\de(x-y) \,\, ,
\no \no
&&
[\bar \psi^a(x), \bar \pi_b(y)]=i \hbar \de^a_{~b} \de(x-y) \,\, ,
\no \no
&&
[\psi^a(x), \bar \pi_b(y)]=[\bar \psi^a(x), \pi_b(y)]=0 \,\, ,
\eq
where $x$ denotes only a space variable, and time is
fixed at a common value. Using the differential representation 
of the momentum implied by the Schr\"odinger representation
\be
\pi_a(x)=-i \hbar {\de \over \de \psi^a(x)} \,\, , 
\Sp \bar \pi_a(x)=-i \hbar {\de \over \de \bar \psi^a(x)},
\ee
the quantum Hamiltonian is given by $(g= N \la)$
\bq
H={g \over N^2} \int dx \Big[-\hbar^2 &&(N+|\psi|^2) 
(N \de^{ab} + \bar \psi^a \psi^b) 
{ \de^2 \over \de \psi^a(x) \de \bar \psi^b(x)}+
\no \no
&&
{N^3 \over g^2} { (N+|\psi|^2) \de_{ab} - 
\bar \psi _a \psi _b \over (N+|\psi|^2)^2} 
\nabla \psi_a \nabla \bar \psi _b \Big] \,\,.
\label{ham1}
\eq
As the $CP(N)$ model generates a  divergent mass
in the quantum level, (see equation (\ref{eq1})
and (\ref{massgap})), 
it is necessary to regularize by introducing a mass counter-term. 
In addition, the 
eigenvalue of the Hamiltonian turns out to diverge. Hence, we 
add the following part to the Hamiltonian
\be
H_{c.t.}={1 \over g} \int dx \tilde \mu _0^2 (N+|\psi|^2) 
+\E_{c.t.} \,\, ,
\label{count}
\ee
where $\tilde \mu _0^2$ and $\E_{c.t.}$ will be determined appropriately 
in the following in order to subtract the divergences of the Hamiltonian.
Note that the  mass counter-term of (\ref{count}) breaks the original
$SU(N+1)$ invariance of $CP(N)$ model, (\ref{actnow}), to its linearly
realized subgroup $SU(N)\times U(1)$.

\section{Variational Method}

The Schr\"odinger equation will be studied within the variational 
approach \cite{jack1,Pi}. Let us take a Gaussian trial wave functional
\be
\Psi[\psi, \bar \psi]= \exp \left\{-\left[ \int dxdy 
(\bar \psi _a(x) - \widehat{\bar \psi}_a (x))
{ G^{-1}_{ab}(x,y) \over 2 \hbar } 
(\psi _b(y) - \widehat{ \psi}_b (y))\right]\right\},
\ee
for some specific configuration $\widehat{\bar \psi}$,  
$\widehat{ \psi}$, and some propagator, $G^{-1}_{ab}(x,y)$, to be determined. 
We suppressed the normalization constant.
Then, the following expectation values result
\bq
&&
 \langle \psi _a (x) \rangle =\widehat{\psi}_a(x) \,\, , 
\Sp  \langle \bar \psi_a(x) \rangle =\widehat{ \bar\psi}_a (x)\,\, , \Sp
 \langle \bar \psi_a(x)\psi_b(y) 
\rangle =\widehat{\bar \psi}_a (x)\widehat{\psi}_b (y)+
\hbar G_{ba}(y,x) \,\, ,
\no \no
&&
 \langle \psi_a(x) \psi_b(y) \rangle =
\widehat{ \psi}_a (x)\widehat{\psi}_b (y) \, , \Ssp
 \langle \bar \psi_a(x) \bar \psi_b(y) \rangle =
\widehat{\bar \psi}_a (x)\widehat{\bar \psi}_b (y) \, ,\Ssp
 \langle \pi_a(x)\bar \pi_b(y) \rangle =
{\hbar \over 4} G^{-1}_{ba}(y,x) \,\, .
\eq
It is not possible to compute the expectation value
of the Hamiltonian,  $ \langle H \rangle $,
in a  closed form,  because of the non-polynomial character of $H$. 
Instead, we take the large $N$ limit (with $g$ fixed), 
and keep the dominant term in $N$. 
In order to simplify the result even more, we take $\widehat{\psi}$ to be 
$x$-independent. Due to the $N$-plicity of the fields in the $CP(N)$ model 
we can set the scales such that $|\hat \psi|^2 \sim N$ 
and $G_{aa}(x,x)\sim N$. 
Hence, we obtain for large $N$ 
\bq
 \langle H \rangle =&&{g \over N} \int dx \Big[ - {\hbar^2  \over 4}
(N+|\hat \psi |^2) (\de(0))^2 + 
{\hbar \over 4}(N+|\hat \psi |^2) G_{aa}^{-1}(x,x) + 
{\hbar ^2 \over 4} G_{aa}(x,x) G^{-1}_{bb} (x,x)
\no \no 
&&
-{ N^2 \hbar \over g^2 (N+|\hat \psi |^2)} 
\left. \nabla^2_x G_{aa}(x,x') 
\right|_{x'=x} +{N^2 \hbar^2 \over g^2 
(N+|\hat \psi |^2)^2} G_{aa}(x,x) \left. 
\nabla^2_x G_{bb}(x,x') \right|_{x'=x}
\no \no
&&
+{N^2 \hbar^2 \over g^2 (N+|\hat \psi |^2)^2} 
\Big(\left. \nabla_x G_{aa}(x,x^\prime) 
\right|_{x'=x}\Big)^2 +{N \over g^2} \tilde 
\mu_0^2 (N+|\hat \psi |^2 +\hbar G_{aa}(x,x)) \Big] +\E_{c.t.}  \,\, ,
\label{ham2}
\eq
where terms are kept only to order $\hbar^2$. In the following we are 
going to keep $\hat \psi $ as an arbitrary parameter, while $G^{-1}$ will
be chosen such that it minimizes the expectation value of the Hamiltonian.
The variation with respect to 
$G_{ab}(x,y)$ gives the following equation
\bq
\Big(1+{ \hbar G_{cc}(x,x) \over N+|\hat \psi |^2} 
&&\Big)G^{-2}_{ab}(x,y)=
\no \no 
&&
{4 \over g^2} {N^2 \over (N+|\hat \psi |^2)^2}
\Big(1-{ \hbar G_{cc}(x,x) \over N+|\hat \psi |^2} 
\Big)\Big( -\nabla_x^2 + m^2\Big) \de(x-y)\de_{ab} \,\, ,
\label{eq1}
\eq
where the finite mass, $m^2$, is given by
\be
m^2 =
{(N+|\hat \psi |^2) \over N} \Big(1-
{ \hbar G_{cc}(x,x) \over N+|\hat \psi |^2}\Big)^{-1} 
\Big( {\hbar g^2 \over 4 N} G^{-1}_{aa} (x,x)+\tilde \mu_0^2\Big)
 \equiv \mu^2+\mu_0^2.
\label{massgap}
\ee

This equation can be solved by performing Fourier transform of 
the delta function, resulting in
\be
G_{ab}^{-1}(x,y)={2 N \over g(N+|\hat \psi|^2)} 
\Big( 1 - {\hbar G_{cc}(x,x) \over N+|\hat \psi|^2} \Big) 
\int {dp \over 2 \pi} \sqrt{p^2 +m^2} e^{ip(x-y)} \de_{ab}
\ee
which yields the mass gap equation: 
\be
\mu^2={g \hbar \over 2} \int {d p\over 2\pi}\sqrt{ p^2 +m^2}.
\label{massgapp}
\ee
Note that the dynamically generated mass $\mu^2$ has a
quadratic divergence (times bare coupling) in terms of the momentum cutoff. 
This is consistent with the one-loop result in perturbation theory
\footnote{ Small $\psi^a, \bar\psi^b$ expansion of the Lagrangian given
in (\ref{actnow}) contains the derivative
interaction of the form $- 1/(gN)(\vert\psi\vert^2\vert
\partial\psi\vert^2+\vert\bar\psi\partial\psi\vert^2)$. This 
lowest order interaction will produce quadratic
divergence in the one-loop order contributing to the 
dynamically generated mass.}.  
Substitution of the variational  equation (\ref{eq1}) into (\ref{ham2})
leads to effective Hamiltonian. We can bring it  into the following
expression;
\be
\langle H \rangle =N[ \int dx] \Big( {m^2 \over g} + 
{\hbar \over 2} \int {dp \over 2 \pi} \sqrt{p^2+m^2} - 
{\hbar m^2 \over 2} \int {dp \over 2 \pi} 
{1 \over \sqrt{p^2 +m^2}} \Big) 
+ \E_{c.t.}
\label{ham4}
\ee
In the above, we only kept terms up to  order $\hbar$ for producing 
the one-loop result. 
In this expression the mass, $m^2$, is finite. However, 
there are infinities connected with the ultra-violet behavior of 
the momentum. In the next section we are going to derive the 
renormalization of the physical quantities of the theory.

\section{Renormalization}

The renormalization of the theory requires that the infinities 
appearing in the Hamiltonian, are absorbed in the bare coupling, $g$, 
the mass, $\tilde \mu^2_0$ and the energy counter-term, $\E_{c.t.}$. 
Introducing a momentum cut-off, $\Lambda$, the finite 
mass, $m^2$, becomes
\be
m^2= \mu_0^2 +\mu^2=\mu_0^2 +{g \over 4 \pi} 
\Big( \Lambda^2+{m^2 \over 2} +m^2 \log {2 \Lambda \over m} \Big)
\label{mas1}
\ee
where the divergences for $\Lambda \to \infty$, have to be absorbed in 
$\mu_0^2$ and $g$. We have set $\hbar=1$. 
In order to treat the infinities, we first define a finite renormalized
coupling constant $g_R$ as
\be
{1\over  g_R}= {1 \over g}+ {1 \over 4 \pi} 
\log {M\over2 \Lambda},
\label{ren1}
\ee
with an arbitrary renormalization scale $M$.
We also define a renormalized mass $\mu_R$ by
\be
\mu_R^2 ={\mu_0^2 +{g \over 4 \pi} \Lambda ^2  \over 1 -  
{g \over 4 \pi} \log {2 \Lambda \over M}}. 
\label{ren2}
\ee
In terms of ({\ref{ren1}) and ({\ref{ren2}), relation 
({\ref{mas1}) becomes
\be
m^2=\mu_R^2 +{g_R m^2  \over 8 \pi} 
\Big(1 + \log {M^2 \over m^2}\Big) \,\, .
\label{renormalized}
\ee
The definitions (\ref{ren1}) and (\ref{ren2})
 can also be viewed as determining 
$\mu_0^2$ and $g$ for finite values of $\mu_R^2$ and $g_R$ with 
respect to the cut-off, $\Lambda$. Eq. (\ref{ren1}) shows
asymptotic freedom as expected.

In terms of the renormalized coupling the Hamiltonian 
in (\ref{ham4}) becomes
\be
\langle H \rangle = N[\int dx] \Big( {m^2 \over g_R} +
{ m^2 \over 8 \pi} \log {m^2 \over M^2} + { m^2 \over 8 \pi} \Big).
\label{finalh}
\ee
where we have chosen the counter-term $\E_{c.t.}$ such that it cancels 
the appearing quadratic divergence, ${1 \over 4 \pi} \Lambda ^2$, of 
the Hamiltonian. This renormalized Hamiltonian is not unique
because the final expression is dependent upon
the renormalization prescription in (\ref{ren1}). 
If one chooses another definition by adding
 a constant $C/4\pi$ term to the right-hand side of
(\ref{ren1}),  the final Hamiltonian (\ref{finalh}) 
changes by $\Delta \langle H\rangle=
N[\int dx](-C m^2/8\pi)$. Note that relation (\ref{renormalized}) 
remains the same. The constant $C$ can be fixed by requiring
a renormalization condition. If we demand, for example, 
\be
{d\langle H \rangle\over d m^2}\Big\vert_{m^2=m^2_0}=0,
\ee
where $m_0^2$ is the solution of (\ref{ren1}),
\be
m_0^2=M^2\exp[-{8\pi\over g_R}],
\ee
we find $C=2$, and this $\langle H\rangle$ agrees with the one 
in the literature \cite{coleman}. Note that $\langle H \rangle |_{m^2=m_0^2}$
does not depend on the coupling constant explicitly, a 
phenomenon known as dimensional transmutation \cite{coleman}.

\section{Conclusions}

We have shown that the variational technique with Gaussian wave 
functional reproduces the known results of $CP(N)$ model in the
large $N$ limit. Unlike the path integral method
which uses an auxiliary field, Hamiltonian procedure necessitates
two counter terms. One is a mass counter-term to treat the divergence
appearing in the mass gap equation and to extract a finite mass. 
The other is a quadratically divergent energy counter-term. 
The remaining divergences in the Hamiltonian can be absorbed by
renormalizing the coupling constant and mass.

The present work can be extended  in a couple of ways. First, 
recall that nonlinear sigma model with target manifold of
symmetric space can also be solved in the large $N$ limit
\cite{pisa}. It would be worthwhile to test the Gaussian
method in this case also. Secondly, it would be interesting to
perform higher-loop analysis which requires a systematic
expansion of Gaussian wave functional \cite{okop}.

\section{Acknowledgment}
We would like to thank Professors R. Jackiw and A. K. Kerman for useful 
discussions. PO is supported by the Korea Science and Engineering 
Foundation  through the project number (95-0702-04-01-3), and
by the ministry of Education through the Research Institute 
for Basic Science(BSRI/ 98-1419).

\end{document}